\documentclass[11pt,twoside]{article}

\usepackage{fancyhdr}
\usepackage{epsfig}
\usepackage{latexsym}
\usepackage{microtype}
\usepackage{lastpage} 
\usepackage{hyperref}
\usepackage{graphicx}
\usepackage{lipsum}
\usepackage{amsthm}
\usepackage{amsmath}
\usepackage{mathrsfs}
\usepackage{yhmath}
\usepackage{amssymb}

\usepackage[dvipsnames*,svgnames]{xcolor}
\hypersetup{colorlinks=true,allcolors=DarkBlue,linkcolor=black}

\newtheorem{definition}{Definition}
\newtheorem{theorem}{Theorem}

\newtheorem{lemma}{Lemma}

\DeclareMathOperator{\dett}{det}
\DeclareMathOperator{\vol}{vol}

\DeclareMathOperator{\spann}{span}
\DeclareMathOperator{\inradius}{inradius}
\DeclareMathOperator{\circumradius}{circumradius}

\newcommand{\trshortyear}{19}
\newcommand{\trpapernumber}{01}
\newcommand{\trmonth}{March}
\newcommand{\tryear}{2019}

\newcommand{\TheAuthor}{}
\newcommand{\Author}[1]{\renewcommand{\TheAuthor}{#1}}

\newcommand{\TheTitle}{}
\newcommand{\Title}[1]{\renewcommand{\TheTitle}{#1}}
\def\R{{\mathbb R}}
\def\NN{{\mathbb N}}
\def\Z{{\mathbb Z}}
\def\x{{\mathbf x}}
\def\y{{\mathbf y}}

\def\v{{\mathbf v}}
\def\u{{\mathbf u}}

\def\zero{{\mathbf 0}}

\Author{Emanuel Florentin OLARIU}
\Title{A note on sequences of lattices}
\newcommand\blfootnote[1]{%
	\begingroup
	\renewcommand\thefootnote{}\footnote{#1}%
	\addtocounter{footnote}{-1}%
	\endgroup
}

\begin{document}

	\blfootnote{The publisher does not claim any copyright for the technical reports. The author keeps the full copyright for the paper, and is thus free to transfer the copyright to a publisher if the paper is accepted for publication elsewhere. }   	

\parindent=8mm

\noindent {{\bf \scriptsize Faculty of Computer Science, Alexandru Ioan Cuza University Ia\c si}}

\noindent {{\bf \scriptsize Technical Report TR \trshortyear-\trpapernumber, \trmonth ~ \tryear}}
\vskip -3mm
\noindent\rule{10.2cm}{0.4pt}
\vskip -1mm
\noindent

\vspace{1cm}
\begin{center}
{\Large\bf A note on sequences of lattices}
\end{center}
\vspace{4mm}

\begin{center}
{\large Emanuel Florentin OLARIU}\footnote{Faculty of Computer Science, Alexandru Ioan Cuza University Ia\c si, General Berthelot 16, 
700483 Ias\c i, Romania, Email: {\tt olariu@info.uaic.ro}}
\end{center}
\vspace{3ex}

\date{}

\begin{abstract}

We investigate the relation between the convergence of a sequence of lattices and the set-theoretic convergence of their corresponding Voronoi cells sequence. We prove that if a sequence of full rank lattices converges to a full rank lattice, then the closures of the limit infimum and limit supremum of the Voronoi cells converges to the corresponding Voronoi cell. It remains an open question if the converse is also true.

\smallskip


\end{abstract}

\section{Introduction}

A {\bf lattice} is a discrete additive group of an euclidean space. Lattices are of great interest for discrete optimization, for example $ \Z^n $, the lattice of points of integral coordinates, is extensively used in integer programming. We analyze the relation between the convergence of a sequence of lattices and the set theoretic convergence of the corresponding Voronoi cells.

We prove that if $ \displaystyle \left( \Lambda_k \right)_{k \ge 1} $ is a sequence of full rank lattices from $ \R^n $ and $ \displaystyle \lim_{k \to \infty} \Lambda_k = \Lambda $ - a full rank lattice also, then
\[ \overline{\liminf_{k \to \infty} V(\Lambda_k)} = \overline{\limsup_{k \to \infty} V(\Lambda_k)} = V(\Lambda), \]
where, for a given sequence $ \displaystyle \left( A_k \right)_{k \ge 1} $ of sets from $ \R^n $
\[ \liminf_{k \to \infty} V(A_k) = \bigcup_{h \ge 1} \bigcap_{j \ge h} V(A_j), \liminf_{k \to \infty} V(A_k) = \bigcap_{h \ge 1} \bigcup_{j \ge h} V(A_j)\]
are the set theoretic limit inferior and limit superior.

\section{Notations and definitions}

For the following definitions one can consult \cite{barvinok11}  and \cite{cassels97}. 

\begin{definition}

A {\bf lattice} $ \Lambda \subset \R^n $ is a discrete additive group of $ \R^n $, i. e.,  $ \x - \y \in \Lambda $, $ \forall\: \x, \y \in \Lambda $ and there exists an $ \epsilon > 0 $, such that $ B(\zero; \epsilon) \cap \Lambda = \varnothing $. Its rank, $ rank(\Lambda) $, is the dimension of its spanned subspace, $ \dim(\spann(\Lambda)) $.

\end{definition}                            

For every lattice there exist $ m $ independent vectors $ \v_1, \v_2, \ldots, \v_m $ in $ \Lambda $, such that 
\[ \Lambda = \left \{ \sum_{i = 1}^m \alpha_i \v_i \: : \: \alpha_1, \ldots, \alpha_m \in \Z \right\}, \]
where $ m = rank(\Lambda) $. This a {\bf basis} of $ \Lambda $. For a given base, if $ A $ is the matrix whose columns are the basis vectors, the {\bf determinant} of $ \Lambda $ is 
\[ \dett(\Lambda) = \sqrt{\dett(A^TA)}. \]

When the lattice has full rank, the determinant is the volume of the {\bf fundamental parallelepiped}:
\[ \Pi = \left\{ \sum_{i = 1}^m a_i\v_i \: : \: a_i \in [0, 1), \forall i = \overline{1, n} \right\}. \]

The fundamental parallelipiped has the following property: every vector $ \x $ in $ lin(\Lambda) $ can be written uniquely as $ \x = \v + \y $, where $ \v \in \Lambda $ and $ \y \in \Pi $. That is the spanned subspace can be {\bf tiled} with copies of the fundamental parallelepiped centered in the vectors of the lattice. The fundamental parallelepiped depends on a particular basis but there is a similar polytope which can be uniquely associated with a lattice:

\begin{definition}

Let $ \Lambda \subseteq \R^n $ be a full rank lattice. Its {\bf Voronoi cell} is $ V(\Lambda) = \{ \x \in \R^n \: : \: \lVert \x \rVert \le \lVert \x - \v \rVert, \forall \: \v \in \Lambda \} $.

\end{definition}

The Voronoi cell has a similar property of $ \spann(\Lambda) $ tessellation:
\[ \spann(\Lambda) = \Lambda + V(\Lambda), \]
where the intersection of different tiles (cells) occurs only on their boundaries (frontiers) and $ \dett(\Lambda) = \vol(V(\Lambda)) $.

The following two lattice parameters (see \cite{cassels97}, \cite{sikiric08}) illustrates the importance of the Voronoi cell. If $ \Lambda $ is a lattice, the {\bf packing radius} of $ \Lambda $, $ \rho(\Lambda) $,  is half the length of the shortest non-zero vector of $ \Lambda $, i. e.
\[ \rho(\Lambda) = (\min_{\v \in \Lambda}{\lVert \v \rVert})/2 = \inradius \left[V(\Lambda) \right]. \]
The {\bf covering radius} of $ \Lambda $ is the smallest $ \mu  $ such that spheres of radius $ \mu $ centered around all the points in $ \Lambda $ cover the entire $ lin(\Lambda) $:
\[ \mu(\Lambda) = \max_{\x \in lin(\Lambda)}{d(\x, \Lambda)} = \circumradius \left[V(\Lambda) \right]. \]

\begin{definition}

 Let $ \Lambda $ and $ \displaystyle \left( \Lambda_k \right)_{k \ge 1} $ be full rank lattices in $ \R^n $, we say that $ \displaystyle \left( \Lambda_k \right)_{k \ge 1} $ {\bf converges} to $ \Lambda $, $ \displaystyle \lim_{k \to \infty} \Lambda_k = \Lambda $ if, for each $ k \ge 1 $, there exists a basis $ \v_{k1}, \v_{k2}, \ldots, \v_{kn} $ of $ \Lambda_k $ such that there exist the limits
\[ \lim_{k \to +\infty} \v_{ki} = \v_i, \forall 1 \le i \le n, \]
and $ \v_1, \v_2, \ldots, \v_n $ is a basis of $ \Lambda $.

\end{definition}

\section{Main result}

\begin{lemma}

Let $ \Lambda \subseteq \R^n $ be a full rank lattice. There exists $ R \in \R^*_+ $ such that $ V(\Lambda) = V_R(\Lambda) = \{ \x \in \R^n \: : \: \lVert \x \rVert \le \lVert \x - \v \rVert|, \forall \: \v \in \Lambda, \lVert \v \rVert \le R \} = \{ \x \in \R^n \: : \: 2\v^T\x \le \lVert \v \rVert^2, \forall \: \v \in \Lambda, \lVert \v \rVert \le R \} $ and $ V(\Lambda) $ is a bounded set. (Consequence: $ V(\Lambda) $ is a polytope.)

\end{lemma}

{\bf proof:} We prove first that $ V(\Lambda) $ is bounded; let $ (\u_i)_{1 \le i \le n} $ be a basis of $ \Lambda $ and $ \displaystyle r = \max_{1 \le i \le n}{\lVert \u_i \rVert} $. If $ \x \in V(\Lambda) $, then $ \displaystyle \x = \sum_{i = 1}^n \beta_i \u_i $, for some $ \beta_i \in \R $, $ \forall 1 \le i \le n $ and
\[ \lVert \x \rVert = \left\lVert \sum_{i = 1}^n [\beta_i] \u_i + \sum_{i = 1}^n \{\beta_i\} \u_i  \right\rVert \le \left\lVert \sum_{i = 1}^n {\beta_i} \u_i  \right\rVert \le \sum_{i = 1}^n \lVert \u_i \rVert \le n \cdot \max_{1 \le i \le n}{\lVert \u_i \rVert} = n \cdot r. \]

Define $ R = 2rn $; obviously, $ V(\Lambda) \subseteq V_R(\Lambda) $. Let $ \x \in V_R(\Lambda) $ and $ \v \in \Lambda $, with $ \lVert \v \rVert \ge R $. We have $ \lVert \x - \v \rVert \ge \lVert \v \rVert - \lVert \x \rVert \ge R - R/2 = R/2 \ge \lVert \x \rVert $. \mbox{\raggedleft{} $ \square $}

\begin{lemma}
Let $ \displaystyle \left( \Lambda_k \right)_{k \ge 1} $ be a sequence of full rank lattices from $ \R^n $. If $ \displaystyle \lim_{k \to \infty} \Lambda_k = \Lambda $, a full rank lattice from $ \R^n $, then there exists $ R' > 0 $ such that $ V(\Lambda) = V_{R'}(\Lambda) $ and $ V(\Lambda_k) = V_{R'}(\Lambda_k) $, $ \forall k \ge 1 $.
\end{lemma}

{\bf proof:} Define $ \displaystyle R_k = 2n \cdot \max_{1 \le i \le n}{\lVert \u_{ki} \rVert} $ and $ \displaystyle R = 2n \cdot \max_{1 \le i \le n}{\lVert \u_{i} \rVert} $; we have $ \displaystyle \lim_{k \to \infty} R_k = R $. There exists $ k_0 \in \NN $ such that $ R_k \le 2R $, $ \forall k \ge k_0 $. Hence by choosing $ R' \ge 2R $ and $ R' \ge R_k $, $ \forall 1 \le k < k_0 $ we have the desired property.
\mbox{\raggedleft{} $ \square $}

\begin{theorem}
\label{cassels97}
(\cite{cassels97}, Theorem 1, V.3) A necessary and sufficient condition that $ \displaystyle  \lim_{k \to +\infty} \Lambda_k = \Lambda $ is that the following two conditions be both satisfied
\begin{itemize}
\item[(i)] if $ \u \in \Lambda $, there are points $ \v_k \in \Lambda_k $, for $ k = 1, 2, \ldots $ such that
\[ \lim_{k \to +\infty} \v_k = \u. \]

\item[(ii)] if $ \x \notin \Lambda $, there is a number $ \epsilon_0 > 0 $ and an integer $ k_0 $, both depending on $ x $, such that
\[ \lVert \u - \x \rVert > \epsilon_0, \forall \: \u \in \Lambda_k, \forall k \ge k_0. \]
\end{itemize}

Where $ \displaystyle \left( \Lambda_k \right)_{k \ge 1} $ and $ \Lambda $ are full rank lattices from $ \R^n $.

\end{theorem}

\begin{theorem}
Let $ \displaystyle \left( \Lambda_k \right)_{k \ge 1} $ be a sequence of full rank lattices from $ \R^n $. If $ \displaystyle \lim_{k \to \infty} \Lambda_k = \Lambda $, a full rank lattice from $ \R^n $, then $ \displaystyle \overline{\liminf_{k \to \infty} V(\Lambda_k)} = \overline{\limsup_{k \to \infty} V(\Lambda_k)} = V(\Lambda) $.
\end{theorem}

{\bf proof:} Suppose that $ (\u_{ki})_{1 \le i \le n} $ is a basis of $ \Lambda_k $, $ \forall k \ge 1 $, and $ (\u_{i})_{1 \le i \le n} $ is a basis of $ \Lambda $ such that $ \displaystyle \lim_{j \to \infty} \u_{ji} = \u_i $, $ \forall 1 \le i \le n $. For every $ \epsilon > 0 $, there exists $ k_{\epsilon} $ such that $ \lVert \u_{ki} - \u_i \rVert \le \epsilon $, $ \forall k \ge k_{\epsilon} $ and $ \forall 1 \le i \le n $. 

Let $ \displaystyle \x \in \liminf_{k \to \infty} V(\Lambda_k) \subseteq \limsup_{k \to \infty} V(\Lambda_k) = \bigcap_{k \ge 1} \bigcup_{j \ge k} V(\Lambda_j) $, i. e., $ \forall k \ge 1 $ there exists $ j = j_k \ge k $ such that $ \x \in V(\Lambda_{j_k}) $. Choose $ \displaystyle \left( \alpha_i \right)_{1 \le i \le n} \subseteq \Z^n $ and $ \displaystyle \u = \sum_{i = 1}^n \alpha_i\u_i \in \Lambda $; we have $ \displaystyle \u_{j_k} = \sum_{i = 1}^n \alpha_i\u_{j_ki} \in \Lambda_{j_k} $ and 

\[ \lVert \x \rVert \le \left \lVert \x - \sum_{i = 1}^n \alpha_i \u_{j_ki} \right\rVert, \forall \: k \ge 1 \stackrel{k \to \infty}{\Longrightarrow} \lVert \x \rVert \le \left \lVert \x - \sum_{i = 1}^n \alpha_i \u_i \right\rVert = \lVert \x - \u \rVert, \]
 hence $ \x \in V(\Lambda) $.

We proved that $ \displaystyle \liminf_{k \to \infty} V(\Lambda_k) \subseteq \limsup_{k \to \infty} V(\Lambda_k) \subseteq V(\Lambda) $. Now we prove that $ V(\Lambda) \subseteq  \displaystyle \overline{\liminf_{k \to \infty} V(\Lambda_k)} $. Suppose not and let $ \x_0 \in V(\Lambda) \setminus \displaystyle \overline{\liminf_{k \to \infty} V(\Lambda_k)} $. It means that there exists an $ \epsilon_0 > 0 $ such that
\[ \displaystyle B(\x_0; \epsilon_0) \cap \displaystyle \liminf_{k \to \infty} V(\Lambda_k) = \varnothing \Leftrightarrow B(\x_0; \epsilon_0) \cap \left( \bigcap_{j \ge k} V(\Lambda_j) \right) = \varnothing, \forall \: k \ge 1.\]
Since $ \x_0 \in V_{R'}(\Lambda) $, we have $ \lVert \x_0 - \u \rVert \ge \lVert \x_0 \rVert $, $ \forall \: \u \in \Lambda $ with $ \lVert \u \rVert \le R' $. By perturbing $ \x_0 $ and decreasing $ \epsilon_0 $ (if necessary), we can suppose that
\[  \x_0 \notin \bigcap_{j \ge k} V(\Lambda_j), \forall \: k \ge 1 \mbox{ and }  \lVert \x_0 - \u \rVert > \lVert \x_0 \rVert, \forall \u \in \Lambda, \lVert \u \rVert \le R'. \]

Using this property we can define an increasing sequence $ \displaystyle (j_k)_{k \ge 1} $ such that $ \x_0 \notin V(\Lambda_{j_k}) $, for any $ k \ge 1 $; hence there exist the points $ \u_{j_k} \in \Lambda_{j_k} $, such that $ \lVert \x_0 - \u_{j_k} \rVert < \lVert \x_0 \rVert $, $ \forall k \ge 1 $. Since $ (\u_{j_k})_{k \ge 1} \subseteq B(\zero; R') $, we can extract a convergent subsequence from $ (\u_{j_k})_{k \ge 1} $, which, for the sake of simplicity, will be named in the same way: $ \displaystyle \lim_{k \to \infty} \u_{j_k} = \u' $. Obviously $ \lVert \x_0 - \u' \rVert \le \lVert \x_0 \rVert $, $ \lVert \u' \rVert \le R' $, and, by (ii) from Theorem \ref{cassels97}, $ \u' \in \Lambda $ - a contradiction with the above property.  \mbox{\raggedleft{} $ \square $}

\section{Conclusions}

The converse of our main result it remains an open problem for now, but it can be proved (which is left for  another technical report) that, if the extremal points of $ V(\Lambda) $ are contained in $\displaystyle \liminf_{k \to \infty} V(\Lambda_k) $ (when this happens, $ V(\Lambda) \subseteq V(\Lambda_k) $, starting from a certain $ k_0 $), then $ \displaystyle \lim_{k \to \infty} \Lambda_k = \Lambda $. It remains to see what happens when some of the extremal points of $ V(\Lambda) $ belong on the boundary of $\displaystyle \liminf_{k \to \infty} V(\Lambda_k) $.

\newpage

\hypertarget{lastpage}{}
\lfoot{{\bf \copyright\hspace{0.01mm} Scientific Annals of Computer Science yyyy}}
\end{document}